# Low-Field Ferroelectric Switching realised by Forced Harmonic Oscillation of Domain Walls

Niyorjyoti Sharma[1,*], Nathan Black[1], Joseph G. M. Guy[1], Eftihia Barnes[2], Kristina M. Holsgrove[1], Brian J. Rodriguez[3,] Raymond G.P. McQuaid[1], J. Marty Gregg[1] and Amit Kumar[1,*]

[1] School of Mathematics and Physics, Queen's University Belfast, Belfast, BT7 1NN, UK

[2] Los Alamos National Laboratory, Los Alamos, NM, 87545, USA

[3] School of Physics, University College Dublin, Belfield, Dublin 4, Ireland

*Address correspondence to: niyor.sharma@qub.ac.uk, a.kumar@qub.ac.uk

**Abstract**

Conventionally, dc fields are used for switching dipole orientations in ferroelectrics. Such fields tilt the potential surface experienced by domain walls and thereby lower activation energies for their movement: escape from tilted potential wells is then realised by thermal excitation, allowing a "creep" process of pinning and depinning to develop. Borrowing ideas of domain wall resonance from the magnetic racetrack community, we show that ac fields, applied at the right frequency, can cause switching at much lower field magnitudes than dc ones (by factors of 4-5). Ferroelectric wall motion appears to be overdamped in the system studied (relaxor strontium barium niobate) and so the maximum in switching efficacy observed, at ~100 kHz, cannot be associated with resonant amplification, which needs an underdamped environment. Instead, in this high viscosity system, the frequency at which the maximum switching efficacy occurs seems to represent a compromise between the attempt frequency for wall depinning (which increases with frequency) and the extent to which energy is transferred to the wall within each field cycle (which decreases with frequency). Notwithstanding the absence of true resonance, the observation that ac excitation can dramatically reduce the bias levels needed for ferroelectric switching could still have significant ramifications for low energy memory technology.

*Introduction*—Growth in digital connectivity has caused Information Communication Technology (ICT) to emerge as a prominent energy-consuming sector worldwide. As per recent estimates, data centres are responsible for 1% of global electricity (>200 TWh) consumption and this is projected to increase to 8% by 2030[1-4]. The carbon footprint of the industry has surged, surpassing even that of the aviation industry[5,6]. Fortunately, aligned with the objective to offset the rising energy demands of data centres, researchers worldwide are exploring alternative, energy-efficient technologies, notably in the realm of low-energy computing and data storage, which constitutes nearly 50% of the total energy consumption in data storage facilities[7].

One of the promising approaches to reduce energy consumption in memory/logic devices would be to transition from widely adopted current-driven memory technologies to their electric field-driven counterparts, as this offers a means to suppress the energy dissipation resulting from Joule heating[8-10]. When discussing electric field-driven devices, ferroelectric materials often emerge as central players. For instance, in the case of ferroelectric memories, the stable spontaneous polarisation state of the material, which defines the memory bits, can be readily written on the application of an electric field. The symmetry of the material preserves the written bit for an extended period without requiring additional energy, hence making ferroelectrics an exceptionally low power-consuming non-volatile memory option. Very recently, electric-field controllability of ferroelectric domain walls has also been exploited to create ferroelectric domain wall logic gates[11,12]. To achieve electric field control over ferromagnetic logic/memory devices, extensive research has been devoted to understanding material systems that demonstrate a direct magneto-electric effect or exhibit indirect coupling between electric field and magnetism through strain[8,9,13,14]. A more complex and highly noteworthy logic architecture was recently proposed by Manipatruni *et al.*, wherein magneto-electric switching in combination with electron spin-orbit coupling was leveraged to achieve lower switching voltage and higher logic density compared to the present-day CMOS technology[15]. They named it 'MESO' logic.

An alternative approach to minimise the energy cost involved in a memory has been demonstrated in magnetic domain wall race-track devices[16]. In this high-density memory architecture, a string of memory bits is stored in a thin film ferromagnetic track, in the form of small magnetic domains. The width of each domain (memory element) is controlled via strategically arranging domain wall pinning sites (often periodically patterned notches in the racetrack). Wall motion is achieved by the application of spin polarised currents, and the associated spin transfer torque that such currents impart, when they encounter a change in magnetic domain orientation. In a landmark publication, Parkin *et al*. demonstrated that the threshold currents required, for domain wall depinning, could be drastically reduced (by a factor of 5) by tunning the duration and the frequency of successive spin polarised current pulses. At the right frequency, sequences of current pulses were found to induce resonant amplification, in which an accumulation of pulses would "rock" domain walls, within their pinning potential wells, to progressively greater energy amplitudes, until they escaped the pinning wells and could hence freely propagate[17,18]. This breakthrough revealed an efficient mechanism of energy transfer to a ferromagnetic domain wall resulting in low-loss magnetisation reversal in racetrack memories. To date, an analogous method for depinning domain walls in ferroelectrics has not been demonstrated, despite its possible technological value.

Here, in this work we show efficient low-field switching in a relaxor ferroelectric material, $Sr_{0.61}Ba_{0.39}Nb_2O_6$ (SBN:61), in a manner strongly reminiscent of the resonant amplification phenomenon discussed above. We find that an AC bias of a chosen frequency induces ferroelectric switching at significantly lower voltage amplitude than a DC bias. Additionally, we observe a frequency dependence in the switching behaviour and a peak in switching efficiencies, implying potential resonant amplification in ferroelectric domain walls. Fitting the frequency dependence to a physical model of a domain wall undergoing damped oscillation within a pinning potential landscape at room temperature revealed an important distinction in comparison to resonance amplification seen in magnetic domain walls. The AC frequency at which the switching efficacy peaks represents a balance between the attempt frequency for domain-wall depinning (increasing with frequency) and the energy transferred to the wall per field cycle (decreasing with frequency). Our work deviates clearly from conventional AC poling strategies where voltages typically higher than coercive voltages are used to induce ferroelectric switching [19-22]. The observation that ac excitation can dramatically reduce the bias levels needed for ferroelectric switching could have significant implications for low energy memory technology.

Strontium Barium Niobate or $Sr_xBa_{1-x}Nb_2O_6$ is a uniaxial ferroelectric system with a tetragonal tungsten bronze structure and is extensively studied for its excellent acousto-optic[23,24], optic[25], nonlinear-optical[26], photorefractive[27], and pyroelectric properties[28]. SBN:100x crystalises to tungsten bronze structure for 0.25<x<0.75, and its ferroelectric properties strongly depend on its composition: Barium-rich SBN exhibits normal ferroelectricity, whereas strontium-rich compositions exhibit relaxor-type ferroelectricity[29]. In all cases, there are only two allowed domain variants which can be easily detected via vertical Piezoresponse Force microscopy (PFM). Hence, the technique has been employed extensively to investigate the labyrinthine domain structure in thermally depolarised SBN single crystals at room temperature as well as to evaluate its domain structure induced by a DC electric field[30-33].

Switching and relaxation dynamics of these domains have also been previously investigated via PFM[34-36]

For this study, investigations were conducted on (001) $Sr_{0.61}Ba_{0.39}Nb_2O_6$ (SBN:61) crystal, obtained from MTI Corporation, with a thickness of ~0.5 mm. This composition sits on the Strontium-rich side of the composition and leans towards relaxor behaviour (**supplementary figure 1**, EDS composition map). Initially, the crystal was heat-treated up to 200 °C for 2 hours to eliminate any residual polarisation in it. After thermal depolarisation, the crystal was cooled back to room temperature and aged for more than a year. Subsequently, we determined the polarisation state of the as-faced sample by poling it with a +/- 100V DC voltage in a "box in a box" geometry using an AFM High-Voltage module. The resulting double box domain microstructure, obtained using vertical PFM, is shown in **supplementary figure 1**. Given that SBN only possesses two domain variants in its ferroelectric state, we conclude that the unpoled areas of the sample have their polarisation oriented along the positive z-axis in the laboratory reference frame. However, because SBN is a relaxor ferroelectric, its unpoled areas may contain nonpolar regions of opposite polarity, although these are not visible in the PFM phase map due to their small size.

*Domain switching with an AC Bias*—After confirming DC bias-induced switching in the crystal, we evaluate AC bias-induced ferroelectric switching in the SBN crystal. An AFM tip, biased with a symmetrical bipolar AC electric field, is scanned over a pristine area on the sample surface, while the sample base is grounded. As shown in **Figure 1** b**,** the AC amplitude must be higher than a threshold value (~2V at 400 µN tip force) for the domains to switch. This implies that PFM performed at a drive amplitude below this threshold can be used for imaging the resultant domain microstructure, post switching. The PFM scans reported in the main text are performed at 1V drive amplitude using a 2.5 N/m conducting tip near the tip-sample contact-resonance frequency of 350 kHz. The domain microstructure remained unaltered even after repeated scanning with 1V drive amplitude, which further supports the feasibility of using low voltage resonance-PFM for evaluating the switched domains.

To further establish the sub-coercive AC-switching behaviour in SBN, we compare the extent of switching at various AC and DC drive amplitudes. Two equivalent 6 µm wide areas were scanned with an AC and a DC tip bias separately, and in each case, the drive amplitude was increased in steps after an interval of 1 µm. **Figure 1**(a, b) compares the resultant domain microstructures obtained during the imaging step using 1V drive amplitude, post DC and AC-switching, respectively. As the drive amplitude increases, the domains coarsen from nano-islands to a fractal structure and ultimately merge into a monodomain state at higher voltages.

This coarsening of the domains can be attributed to the lateral growth of individual domains, driven by the increased energy input associated with higher drive amplitudes.

The fractal domain structure of SBN necessitates the implementation of statistical methods for quantifying the extent of switching at each drive amplitude[32,37]. As demonstrated in **Figure 1** c (and **supplementary figure 2**), we have quantified the extent of switching as "percentage switched": the ratio of the number of pixels switched to the total number of pixels in a given area. A monotonically increasing trend between the percentage switched and the drive amplitude is apparent in both the AC and DC cases. While a 4V DC bias fails to induce any discernible polarisation reversal, an AC bias of identical magnitude switches 70-80% of the scanned area. Moreover, in the case of DC-switching, the onset of polarisation reversal occurs at voltages exceeding 10 V, which is ~5 times higher than what is required to initiate AC-switching. Measurements with a higher stiffness 80 N/m conducting diamond tip with a larger tip-sample contact area, reveal a similar trend where switching with an AC bias is more effective than DC (**supplementary figure 3**). These results illustrate that the sub-coercive domain switching achieved by an AC scanning bias is fundamentally different from conventional DC switching.

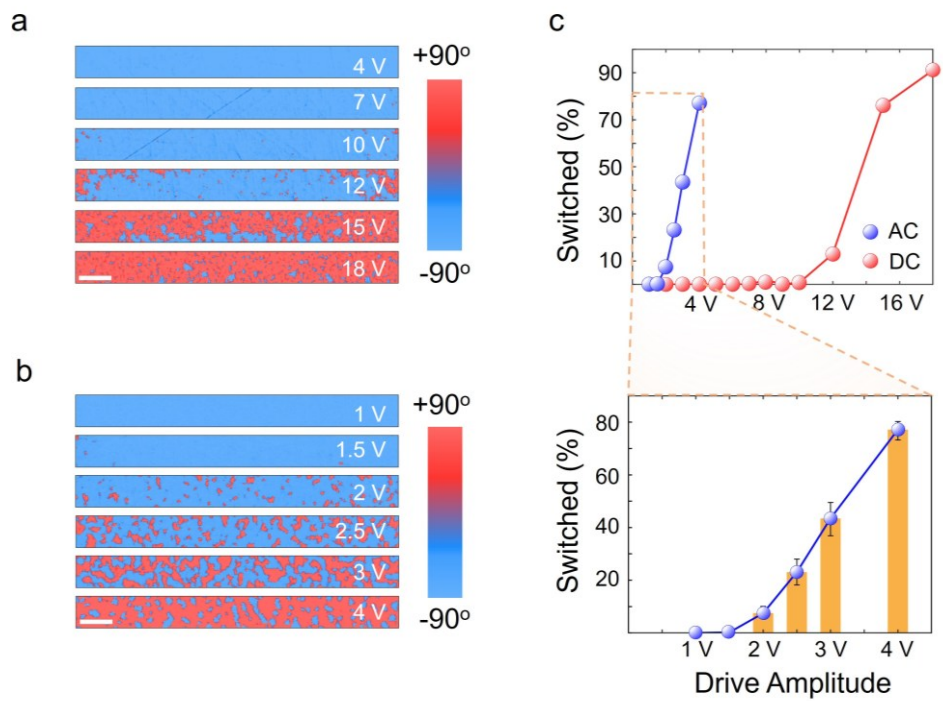

**Figure**1: AC vs DC bias mediated switching: (a) PFM phase obtained post DC switching at voltages ranging from 4 V to 18 V (b) PFM phase maps obtained post AC switching at voltages ranging from 1V to 4V with a frequency of 20 kHz (c) Direct comparison of the efficacies of AC and DC bias on domain switching, drawn statistically though percentage switched. All indicated scale bars are 600 nm.

To gain further insights into the switching dynamics under an oscillating electric field, we investigate into the extent of switching in the frequency space. Switching scans were performed at multiple areas of the sample at varied AC frequencies ranging from 2 Hz to 1 MHz, while keeping the drive amplitude constant at 3V. The PFM phase images obtained subsequently are shown in **Figure 2**. Strikingly, we found the highest percentage switching in a specific frequency range (between 20-200 kHz), resembling a resonance effect. It is important to mention that we kept the scanning speed constant across all measurements. This ensured that the tip spends an equal amount of time at each pixel, hence eliminating the possibility of increased switching due to prolonged exposure to the switching field[37-39]. However, a constant tip speed also implies that the number of AC cycles experienced by each pixel increases proportionally with the applied AC frequency. If the number of AC cycles were the sole determinant of switching efficacy, one would anticipate a monotonic increase in switching with increasing drive frequency. The observed resonant behaviour, therefore, indicates the presence of additional factors influencing the AC switching dynamics, perhaps related to how the domain walls react to different AC frequencies. It is worth noting here that the observed resonance around 20-200 kHz cannot be correlated to the contact resonance of the tip and the sample since the typical contact resonance for the 2.5 N/m AFM tip is around 350 kHz. A similar trend in the switching behaviour, characterised by a peak around 20 kHz, is observed in response to the applied AC frequency when tested with an 80 N/m diamond tip, which has its contact resonance around 1 MHz (**supplementary figure 4**).

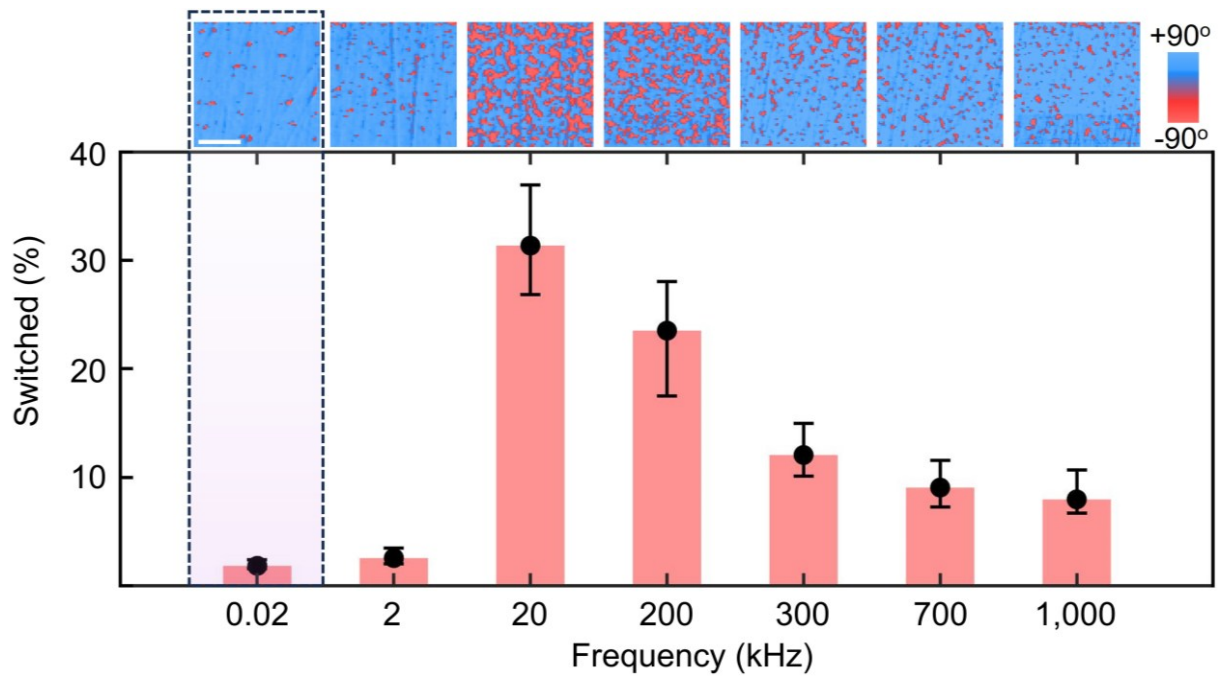

**Figure** 2: Effect of AC frequency on switching efficiency. The top section displays PFM phase maps acquired after switching with a 3V AC bias at various frequencies (as indicated on the x-axis), while the bar chart below quantifies the degree of switching resulting from the AC bias applied at the chosen frequency. The scale bar for each phase map is 1 µm.

*Effect of tip stress and flexoelectricity*—Before discussing the domain wall dynamics in SBN under an applied AC field, it is imperative to address the question: how does a symmetrical bipolar electric field induce stable ferroelectric switching in the first place? An additional uniaxial field should exist, which perturbs the equivalence between the polarisation states in terms of their response to the symmetrical AC field. To verify the existence of such a uniaxial field, we investigated the polarity of AC-induced switching. For this, a 'box-in-a-box' domain microstructure was created over a pristine sample surface, as shown in **Figure 3** b, by scanning an AFM tip biased with a +/-50V. Within each DC bias written region, two small areas (highlighted as white and black dashed boxes) were further scanned with a 3V AC bias. The AC bias only switched the upward pointing written domain, confirming the unipolar nature of AC-switching in the shown experiments and the existence of a uniaxial field. Additionally, the result suggests that the observed AC switching is not limited to the unpoled regions of the sample, where nano-polar regions are expected, but also extends to poled regions, where the nucleation of nano-domains must precede the subsequent growth step.

We contend that the underpinning uniaxial field facilitating AC bias induced unidirectional switching is the flexoelectric field, induced by the AFM probe on the sample surface while in contact (**Figure 2** a)[40-43]. In this context, we investigate the effect of tip stress on AC switching and find that higher tip forces indeed decrease the threshold drive amplitude for AC switching. As **Figure 2** c illustrates, a 3V AC drive amplitude at a tip force of ~200 nN is required to initiate switching, whereas the same effect can be achieved with a 2V AC drive amplitude at a tip force of ~400 nN. Similarly, a 2V AC bias with a tip force of ~1400 nN yields the same effect as a 3V AC bias with a tip force of ~300 nN, as shown in **supplementary figure 5.** These results highlight the key role played by the flexoelectric field in enabling AC switching by introducing asymmetry to the Landau potential landscape, which is expected to be a symmetric double well for a uniaxial ferroelectric in the absence of a DC field.

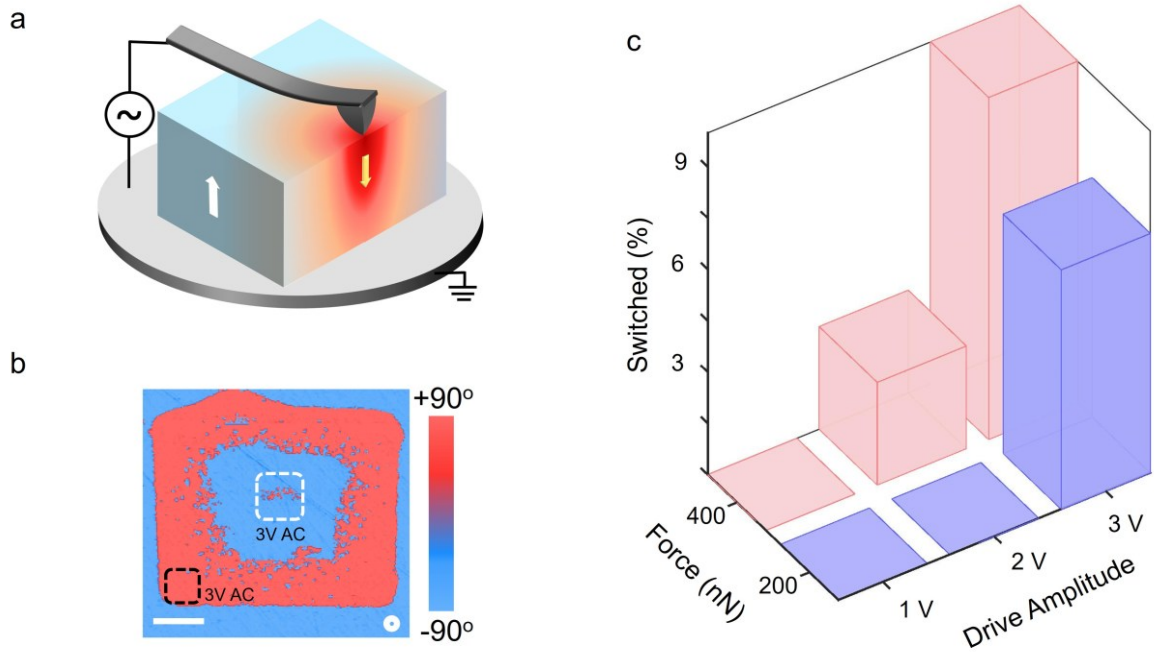

**Figure** 3: Effect of tip stress on AC switching. (a) Schematic illustrating how the flexoelectric field, depicted by a yellow arrow, opposes the bulk polarisation, indicated by the white arrow, of SBN, and thus assisting in up-to-down polarisation switching. (b) AC bias-induced domain writing on regions poled upward (white dotted box) and downward (black dotted box) illustrates the unipolar nature of AC switching. The 'box-in-box' domain structure was written by applying a DC bias of +/- 50V, and the AC switching within them was achieved by applying a 3V AC bias at 330 kHz. The scale bar is 2.25 μm. (c) Extent of switching upon application of an AC bias ranging from 1V to 3V at two different tip stress levels: 200 nN and 400 nN. The drive frequency was set to 330 kHz during both the writing and imaging steps. A 200 nN tip force equates to a 1 V deflection set point.

*Domain wall dynamics in an oscillating field under thermal influence*—The flexoelectric field justifies the asymmetry in the double well potential landscape, but it does not fully answer how the AC bias maximises switching within a specific frequency range at lower voltages than the corresponding DC bias. This led us to consider the dynamics of ferroelectric domain walls in SBN, assuming they (i) possess an effective mass; (ii) undergo damped harmonic oscillations at pinning sites; and (iii) are influenced by both electrical (AC/DC) and thermal energy at room temperature. Kittel first treated a ferroelectric domain wall as a body with an effective mass and derived its equation of motion in a damped medium under an electric field[44] . For an oscillating electric field of frequency ω, the equation of motion of a domain wall of effective $m$, is given by:

$$m\frac{d^2x}{dt^2} + b\frac{dx}{dt} + kx = F_o \sin \omega t \qquad (1)$$

where, $b$ is the damping constant, $-kx$ is the restoring force and $F_o$ is the maximum driving force linked to the drive amplitude. Kittel proposed that the damping constant, $b$ (causing sluggish dipolar rotation during switching), originates from factors like selective impurity diffusion, acoustic radiation, or coupling with lattice vibrations, while the restoring force, $-kx$, results from local trapping of the domain wall at a defect site within the material. The assumption of harmonic domain wall motion (or the defect potential well is parabolic) aligns with that used for describing resonant amplification in magnetic domain walls[16,17].

In relaxor SBN, domain walls are regions of polar discontinuity separating nano-polar regions from their surroundings. Under an applied field, these walls attempt to overcome the pinning sites (which initially stabilised the nano-domains), before merging with adjacent nano-domains to form the characteristic fractal domain microstructure. The extent of coarsening in the domain microstructure—represented by the percentage of switched domains or switching probability—reflects the effectiveness of domain walls depinning under the applied field conditions. This implies that the switching probabilities obtained experimentally must be a function of domain-wall energy at a given AC field amplitude and frequency. The energy of an oscillating domain wall excited by an oscillating bias of frequency $\omega$ is given by:

$$E_o(\omega) = \frac{\omega_0^2 \frac{F_o^2}{2m}}{\beta^2 \omega^2 + (\omega_0^2 - \omega^2)^2} \qquad (2)$$

where $\omega_0$ is the natural frequency of the domain wall ($\sqrt{\frac{k}{m}}$) and $\beta$ is the effective damping constant ($\frac{b}{m}$) of the relaxor medium.

In addition to the electric field, thermal excitations also contribute to the total energy input and influence the dynamics of a domain wall in both ferroelectric[38,45-48] and ferromagnetic[49] systems. Since AC switching in SBN occurs at a sub-coercive voltage, the vibrational energy of a domain wall (under an AC excitation) is expected to be less than the total activation energy necessary to overcome the pinning potential well. Thermal contributions can fulfil this energy deficit; therefore, we equate thermal energy to the difference between the activation energy ($E_a$) and the vibrational energy of the domain wall $E_o(\omega)$. At temperature $T$, the probability of the domain wall possessing the required thermal energy for overcoming the activation barrier ($E_a$), as derived from Boltzmann statistics is given by:

$$p(\omega) = \exp\left(-\frac{E_a - E_o(\omega)}{k_b T}\right) = \exp\left(-\frac{E_a}{k_b T}\left(1 - \frac{\omega_0^2 \frac{F_o^2}{2E_a * m}}{\beta^2 \omega^2 + (\omega_0^2 - \omega^2)^2}\right)\right) \qquad (3)$$

It is worth noting here that the above equation expresses the probability of a domain wall making a single attempt to escape a pinning site. This probability will accumulate as the domain wall attempts to overcome the activation barrier over multiple AC cycles. Given that we use a scanning AC bias to switch SBN, we define the number of attempts a domain wall makes to escape a pinning site as the number of AC cycles experienced by each pixel in a 256 × 256 scanning grid. For an AC frequency $\omega$ and a scan rate of 0.5 Hz, the number of attempts, $n$, is given by $\frac{\omega}{256}$. Hence, the cumulative probability of a domain wall overcoming the activation barrier in $n$ attempts, resulting in AC switching is:

$$P(\omega) = 1 - \big(1 - p(\omega)\big)^n$$

For $p(\omega) \ll 1$,

$$P(\omega) = \left(\frac{\omega}{256}\right) \exp\left\{-\frac{E_a}{k_b T}\left(1 - \frac{\delta}{\beta^2\omega^2+(\omega_0^2-\omega^2)^2}\right)\right\}; \quad \delta = \frac{\omega_0^2 F_o^2}{2E_a * m} \qquad (4)$$

The experimental observations of switching probabilities as a function of AC frequency (**Figure 2**) have been compared to the model of a domain wall attempting to escape a pinning site under the influence of electrical and thermal energy. The key parameters affecting the switching probabilities in equation 4 are the natural frequency of the domain wall ($\omega_0$), the effective damping constant ($\beta$) and the activation barrier ($E_a$). Since the drive amplitude was kept constant during the frequency experiments, $F_o$ and thereby $\delta$ is also treated as constant. The value of $\delta$ is devoid of a physical meaning by itself, as it contains two unknowns: $F_o$ and $m$. We therefore treat $\delta$ as an arbitrary constant during fitting. Nevertheless, the quantity $\frac{\delta}{\beta^2\omega^2+(\omega_0^2-\omega^2)^2}$ represents $\frac{E_o}{E_a}$, and provides the fraction of the activation energy achieved by the domain wall due to its oscillations under an AC field.

A pertinent question arises: why does domain wall relaxation (characterised by a drop in the switching probabilities) occur at a considerably lower frequency than typical dipolar relaxation, which occurs in the gigahertz range? For instance, in the case of Barium titanate, the ferroelectric domain boundaries are predicted to resonate at an AC frequency of 2 GHz in a damping-free medium[44] and the same manifests experimentally as dielectric relaxation around the same frequency range[50]. Permittivity measurements conducted on relaxor SBN:81, below ferroelectric transition temperature (~330 K), also reveal excitations near 10 GHz, attributed to domain wall oscillations[51]. We, therefore, treat the resonant frequency of ferroelectric domain walls as an inherent property, arising from dipolar relaxations, and assign the natural frequency of ~2 GHz (within the correct order of magnitude) to the domain wall, consistent with Kittel's prediction. We fit equation 4 to the frequency dependence of AC switching and extract the remaining parameters: the activation barrier ($E_a$) and the effective damping constant (β). The optimal fit was obtained at an $E_a$ of ~10 $k_bT$ or 0.2 $eV$ (at $T = 300K$), consistent with literature values[52] and a β value of $10^{13}\ s^{-1}$ as illustrated in **figure 4** b.

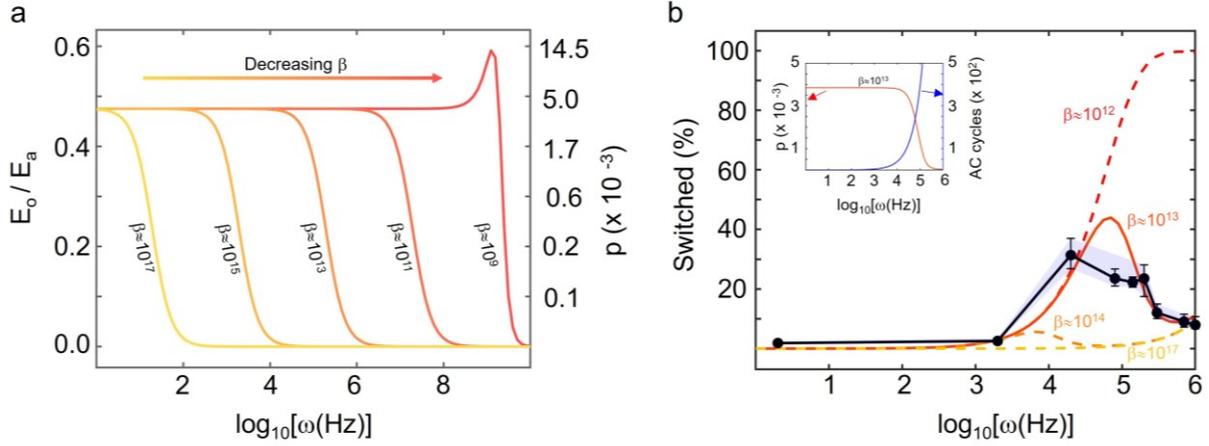

**Figure** 4: Predicted model of an AC bias driven domain wall in a pinned potential landscape. a) Dependence of the ratio of the domain wall's vibrational energy to the activation barrier (left axis) and the probability of switching in a single attempt (right axis) on the drive frequency of the applied AC field in different damping mediums, where β = b/m. (b) Model predictions for the extent of switching and AC frequency at varied effective damping constants are plotted alongside experimental data, showing the best fit at β ~$10^{13}$, depicted by the solid line. The inset in panel b shows the dependence of the probability of switching in one attempt on the applied AC frequency at β ~ $10^{13}$ (left axis) and the AC-cycles or the number of attempts made by the domain wall at a given frequency (right axis).

The fit reveals that domain wall motion in SBN is overdamped, which explains the relaxation of switching probabilities at frequencies four orders of magnitude lower than typical dipolar relaxations. To further illustrate this, we plot the vibrational energy of the domain wall (normalised to the activation energy, i.e. $\frac{E_o}{E_a}$) against AC-frequency, under varying effective damping constants, at a chosen driving frequency of 2 GHz (**Figure 4 a**). For lower values of β (e.g. $10^9\ s^{-1}$ reported for Barium titanate domain walls[53]) domain wall relaxation occurs in the GHz range, whereas for higher values of β (around $10^{13}s^{-1}$), it shifts to lower frequencies ranges : 0.1-1 MHz as observed in our case. A higher value of β in SBN, compared to a normal ferroelectric like Barium Titanate, is not surprising. This is due to inhomogeneities created by random field effects in a relaxor medium, which can lead to a higher value of the damping constant ($b$).

In the low-frequency regime, the vibrational energy ($E_o/E_a$) fails to explain the monotonic increase in the switching probabilities up to a frequency of ~20 kHz. Instead, thermal contributions account for the trend, as the probability of a domain wall overcoming the activation barrier, via thermal excitation, rises with successive attempts during increasing AC cycles. This explains why an AC bias outperforms a DC bias of comparable magnitudes in inducing ferroelectric reversal. One can thus treat DC bias as a limiting case of an extremely low-frequency AC bias. Although the single-attempt switching probability (given by $E_o/E_a$) of a DC bias equals that of a 20 kHz AC bias, the cumulative switching probability under AC excitation is substantially higher, with ~ 80 attempts compared to the DC bias's single attempt within the same time frame.

The model provides a good fit to the experimental data on the switching probabilities as a function of AC frequency; however there are a few limitations to the model that are worth highlighting. The model assumes the motion of the domain wall to be harmonic, suggesting the potential for expansion by incorporating anharmonic components into the equation of motion[54]. It also suggests that even at very low domain wall amplitudes ($\frac{E_o}{E_a} \approx 0$), the domain wall depinning can occur if the number of attempts increases proportionately: for example, by increasing the AC frequency or the dwell time of the AFM tip. However, as **Figure 1** shows, the AC voltage must exceed a threshold (2 V, in this case) for switching to occur, and repeated scanning below this threshold bias did not cause further switching during the imaging step. This implies that there may be a characteristic time within which the domain wall system has to be excited for it to escape a pinning site. It is also possible that only a certain (maximum) percentage of the total energy required for depinning can be achieved thermally. These considerations present clear pathways for future model development and experiments. Nevertheless, the model presented here provides a robust framework wherein domain wall dynamics emerge as a key factor in describing ferroelectric switching under an oscillating bias.

In summary, the observed resonance-like behaviour in SBN—characterised by a peak in the switching efficiency at ~20-200 kHz AC bias—is attributed to two competing factors. First, the number of attempts, which is directly proportional to the AC frequency, explains the observed rise in switching efficiencies at lower frequencies. Second, the vibrational energy of the domain wall is gained from an oscillating electric field. Above a certain frequency, determined by the natural frequency and the damping constant, the domain wall fails to respond to the rapidly changing electric field, resulting in a relaxation of its switching efficiencies. This compromise between the number of attempts ($n$) and domain wall energy due to oscillations ($E_o$) is depicted in the inset **figure 4** b. Clearly, the resonant-like behaviour observed in AC switching in relaxor SBN is different from the true resonance observed in magnetic domain walls. In the latter case, the energy acquired by the domain wall due to its oscillation, indicated by its amplitude, solely governs the probability of switching at a given frequency [51]. At resonance, the maximum amplitude of domain wall vibration leads to maximum energy, allowing the magnetic domain wall to escape the pinning potential.

*Conclusion*—We have demonstrated that an oscillating electric field facilitates more efficient unipolar ferroelectric switching in SBN relaxor crystals compared to a static field. A switching AC bias results in a labyrinthine domain pattern that coarsens with increasing drive amplitude. Localised stress, through an AFM tip, reduces the threshold bias for AC switching and enhances switching efficiencies. The AC switching behaviour further depends on the frequency of the applied AC bias and this can be explained from domain wall dynamics in a parabolic potential well under an oscillating electric field and thermal energy. The results reveal the intricate physics underlying the AC switching behaviour in relaxor ferroelectrics, which resembles 'resonance amplification' in magnetic domain walls, yet is very distinct from one another. The results offer an alternative approach to domain switching in relaxor ferroelectrics with reduced applied bias, which could have implications for SBN-based electronic devices with low operating voltage.

# Low-Field Switching of Ferroelectrics Realised by Forced Harmonic Oscillation of Domain Walls

*Supplementary Information for main manuscript*

Niyorjyoti Sharma,[1,*] Nathan Black,[1] Joseph G. M. Guy,[1] Eftihia Barnes,[2] Kristina M. Holsgrove,[1] Brian J. Rodriguez,[3] Raymond G.P. McQuaid,[1] J. Marty Gregg,[1] and Amit Kumar,[1,*]

[1] School of Mathematics and Physics, Queen's University Belfast, Belfast, BT7 1NN, UK

[2] Los Alamos National Laboratory, Los Alamos, NM, 87545, USA

[3] School of Physics, University College Dublin, Belfield, Dublin 4, Ireland

*Address correspondence to: niyor.sharma@qub.ac.uk, a.kumar@qub.ac.uk

This document presents additional experiments/descriptions that support the findings presented in the main text. These include: 1) verifying the composition of the SBN sample 2) determining the polar orientation of the scanned face 3) describing the methodology implemented for quantifying the extent of switching 4) switching with a 80 N/m stiff AFM tip at varied electric field amplitudes and frequencies 5) additional experiments showing the influence of tip stress on AC switching.

## 1) Determining the composition and the polarisation state of as-found SBN:

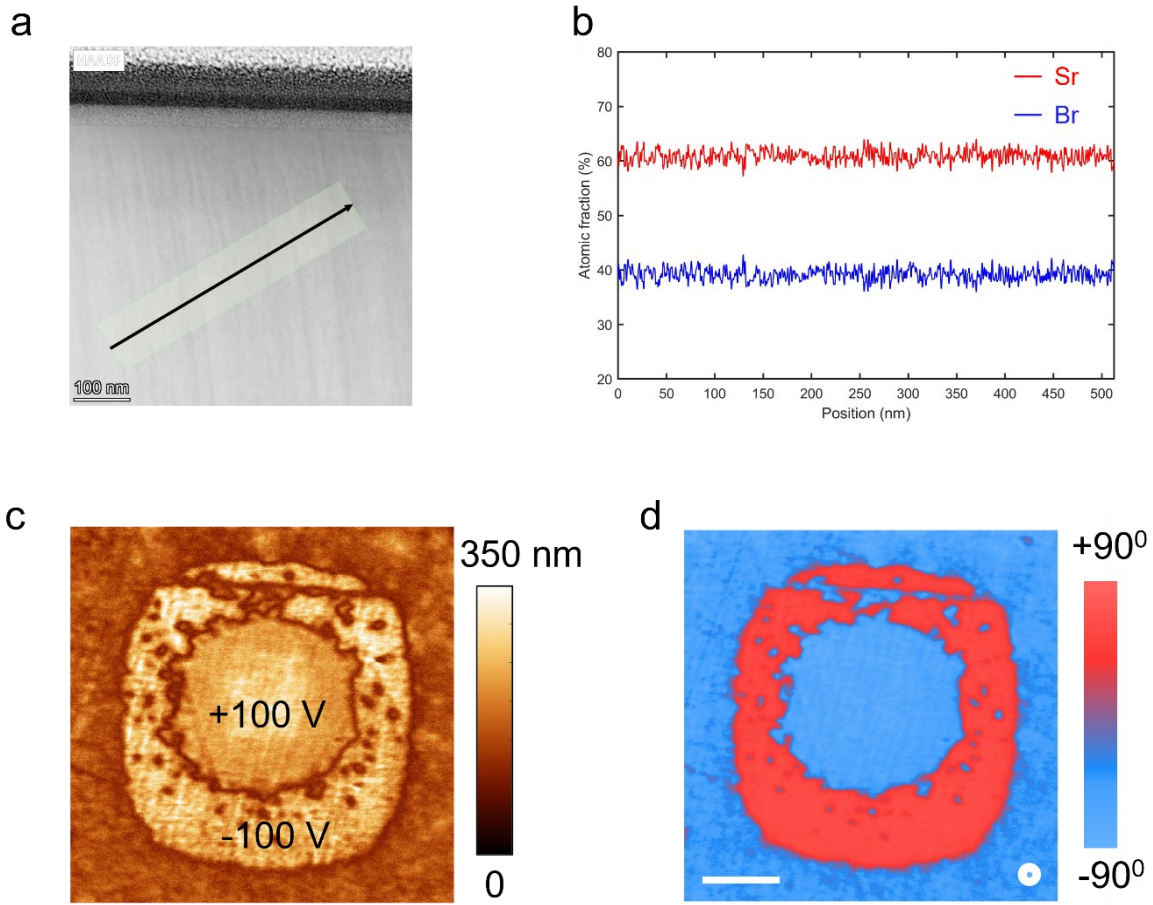

**Supplementary figure** 1: (a) HAADF-STEM image of a lamella cut from the SBN:61 crystal. (b) The EDS line profile along the black line in (a) shows the atomic fraction of Sr (61%) and Ba (39%) in the studied SBN crystal. The X-ray emission lines used were Sr K-α and Ba L-α. (c) PFM amplitude and (d) phase were obtained after switching with a ±100 V DC bias, illustrating the initial polarity of the sample (+z in the laboratory axis). The scale bar for (c) and (d) is 1.8 µm. TEM samples were prepared using a Tescan focused ion beam (FIB)-secondary electron microscope (Lyra 3), where the lamella was sequentially thinned with 30 kV, 15 kV, and 5 kV ion beams. High-angle annular dark-field scanning transmission electron microscopy (HAADF-STEM) was performed on a Thermo Fisher Scientific Talos F200X operating at 200 kV.

## 2) AC vs DC complete image with binarisation:

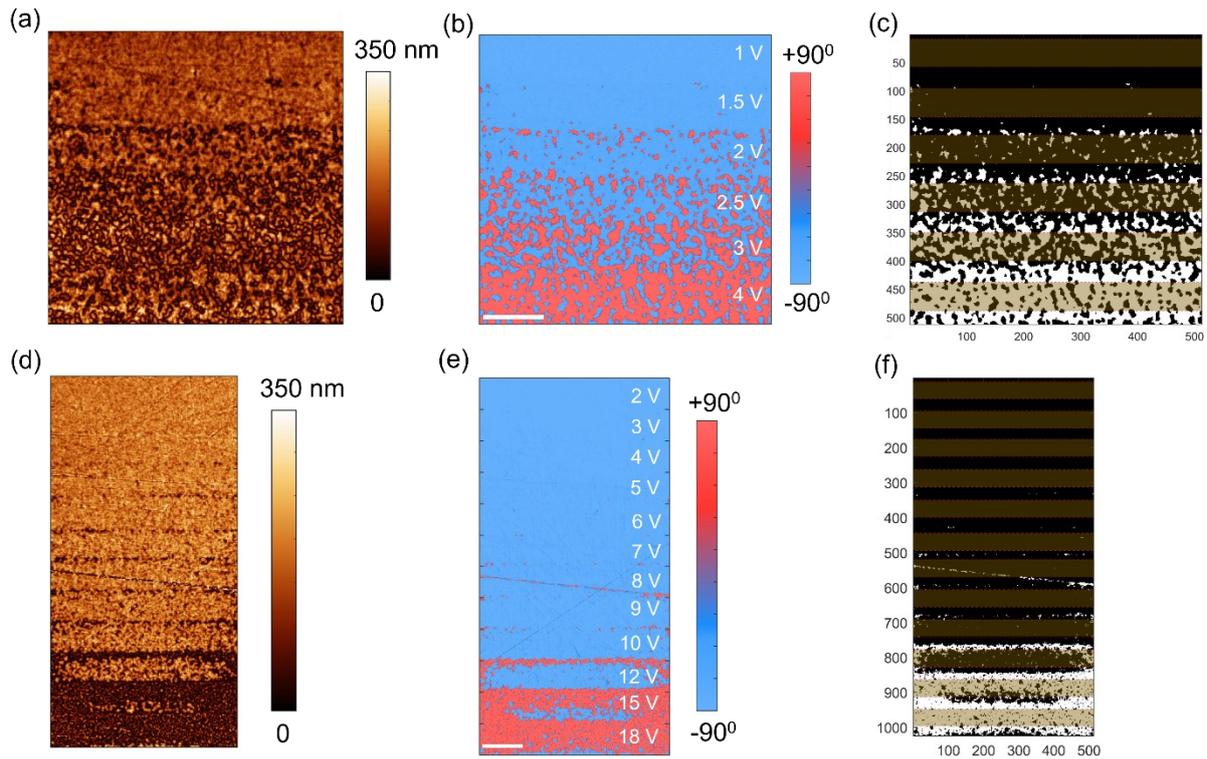

**Supplementary figure** 2: (a) PFM amplitude, (b) PFM phase, and (c) binarised phase image of the area after AC switching with drive amplitudes mentioned in (b). (d) PFM amplitude, (e) PFM phase, and (f) binarised phase image of the area after DC switching, respectively. The highlighted rectangles in (c) and (f) illustrate the cropped locations within which the switching percentages are calculated (for Figure 1 in the main text), avoiding areas of sudden voltage jumps. All scale bars are 1.2 µm. The phase images are converted to a binary image by manually setting a global threshold, and the switching percentages are given by the ratio between the number of pixels appearing bright in the binarised phase map and the total number of pixels within that area. The global threshold lies within a small range instead of being an exact number; hence, the switching percentages calculated at the minimum and the maximum of this range (determined manually) give the error bars in all figures in the main text.

### 3) AC-switching with a 80 N/m diamond coated tip:

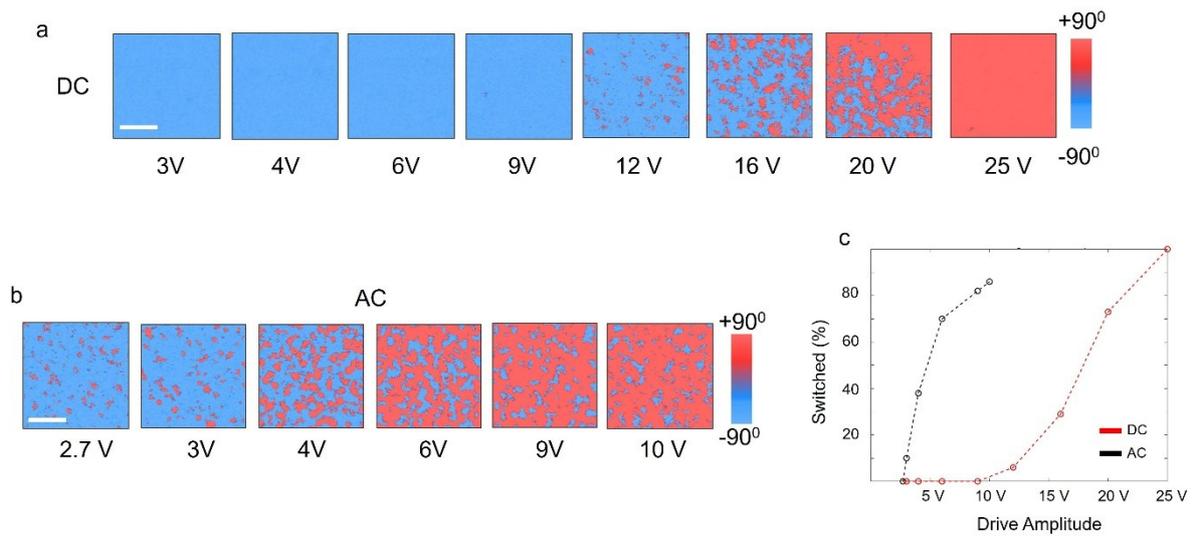

**Supplementary figure** 3: Comparison between AC and DC bias switching using a 80 N/m diamond-coated tip: Phase maps acquired after (a) DC and (b) AC switching (20 kHz) at various drive amplitudes. The deflection setpoint was set to 0.5V during the scans. Scale bar in all panels is 1 μm.

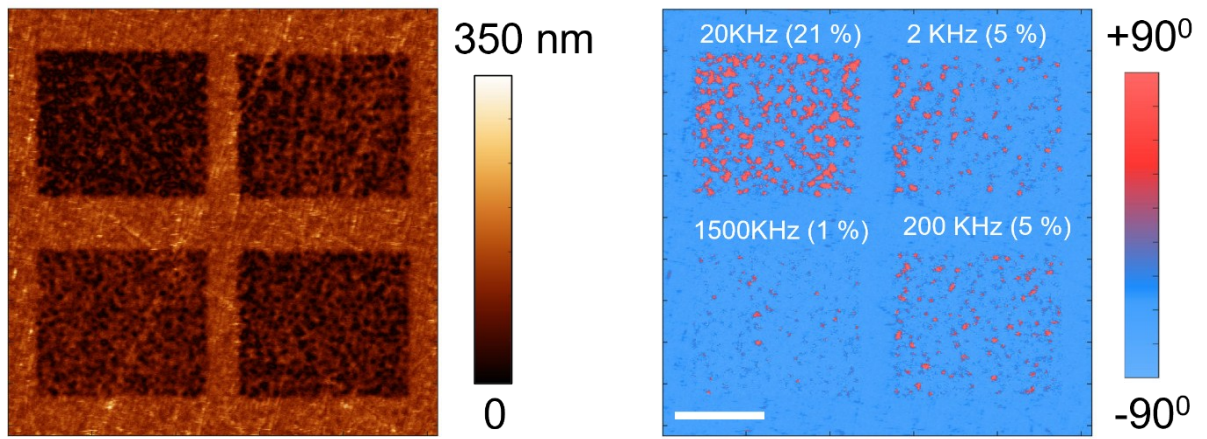

**Supplementary figure** 4: Consistency in the relationship between AC switching and the drive frequency despite variations in tip-sample contact resonance: Amplitude (a) and Phase(b) post AC switching performed with an 80 N/m diamond tip (contact resonance at 1 MHz) at varied frequencies while keeping the drive amplitude at 3V. The scale bar is 3 μm.

4) **Additional experiments demonstrating the effect of tip stress on AC switching**

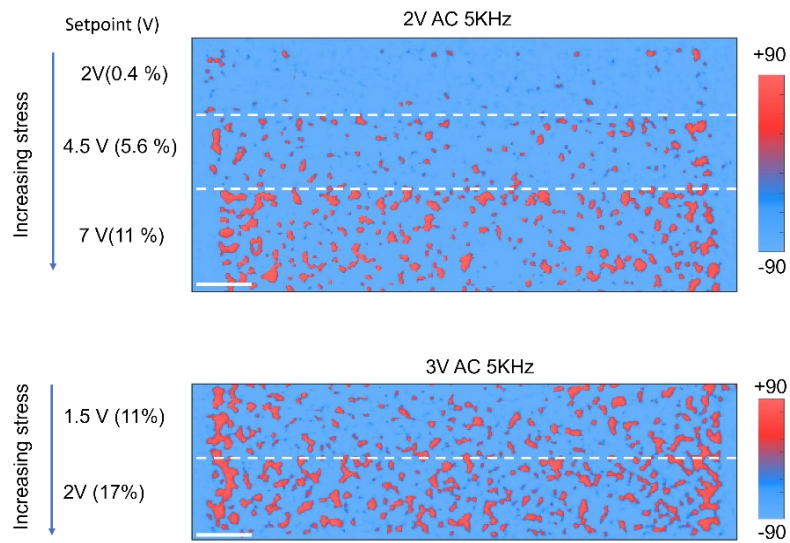

**Supplementary figure** 5:. AC switching performed with a 2.5 N/m PPP-EFM tip at varied tip stress (deflection setpoint) and varied drive amplitude (2V and 3V) while keeping the frequency fixed at 5 kHz. The scale bar is 0.75 µm